\documentclass[amsmath,amssymb,pra]{revtex4}
\usepackage{graphicx}

\usepackage{amsmath}
\usepackage{amssymb}
\usepackage{amsfonts}
\usepackage{color}
\usepackage{float}
\newcommand{\ave}[1]{\langle{#1}\rangle}
\newcommand{\phys}{{\sf P}}
\newcommand{\tape}{{\sf M}}

\begin{document}
\title{Operational derivation of Boltzmann distribution with Maxwell's demon model}
\author{Akio Hosoya$^1$, Koji Maruyama$^{2}$, Yutaka Shikano$^{3,4,5}$}

\affiliation{$^{1}$Department of Physics, Tokyo Institute of Technology, Tokyo, 152-8551 Japan}

\affiliation{$^{2}$Department of Chemistry and Materials Science, Osaka City University, Osaka,
558-8585 Japan}

\affiliation{$^{3}$Research Center of Integrative Molecular Systems (CIMoS), Institute for
Molecular Science, Okazaki, 444-8585 Japan}

\affiliation{$^{4}$Institute for Quantum Studies, Chapman University, Orange, CA 92866, USA}

\affiliation{$^5$ Materials and Structures Laboratory, Tokyo Institute of Technology, Yokohama,
Kanagawa, 226-8503 Japan}

%\date{\today}

\begin{abstract}
The resolution of the Maxwell's demon paradox linked thermodynamics with information theory
through information erasure principle. By considering a demon endowed with a Turing-machine
consisting of a memory tape and a processor, we attempt to explore the link towards the
foundations of statistical mechanics and to derive results therein in an \textit{operational}
manner. Here, we present a derivation of the Boltzmann distribution in equilibrium as an example,
without hypothesizing the principle of maximum entropy. Further, since the model can be applied to
non-equilibrium processes, in principle, we demonstrate the dissipation-fluctuation relation to
show the possibility in this direction.

\end{abstract}

\maketitle

%%%%%%%%%%%%%%%%%%%%%%%%%%%
\section*{Introduction}
%%%%%%%%%%%%%%%%%%%%%%%%%%%
Statistical mechanics has been developed in order to describe the behavior of systems that have a
large number of microscopic degrees of freedom so that it is consistent with thermodynamics
\cite{gibbs1902}. While it is no doubt the best theory we have today to explain the dynamics of
such systems, its foundations are not as solid as they may appear. Particularly, the principle of
equal \textit{a priori} probabilities, or the ergodicity of the system, lacks a clear physical
rationale, which led to coexistence of various approaches on which the theory is
based~\cite{tolman2010,terHaar1955,landau_lifshitz,Khinchin1949}. The history of each school can
be found in e.g., Ref\cite{uffink2006}, and also in the references of a more recent research
paper\cite{reimann2008}. This situation is not very comfortable also from the standpoint that
physical laws should be constructed based on physical operations, even in a thought experiment, as
in the Newtonian mechanics, electromagnetism, and the theory of special relativity.

Thermodynamics, on the other hand, is constructed upon firmly established empirical and
operational evidence on macroscopic objects~\cite{lieb1999}. Further, it is believed to explain a variety of
physical phenomena, regardless of the details of the system constituents. Thus we take the
universality and robustness of thermodynamics as a guiding principle in our attempt to lay the
foundations of statistical mechanics~\cite{chandler1987,hill1987}.

Our motivation is in describing physics in terms of operations, i.e., under the concept of
operationalism~\cite{bridgman,brillouin,brillouin_book,Shikano}. In this respect, we need the notion of probability in the consideration to bridge
thermodynamics and statistical mechanics and it should be introduced through operations.
Fortunately, from the viewpoint of the frequentism~\cite{kendall}, probabilities can be defined as a limit of
relative frequencies of events in a large number of trials or operations. Then, the standard
information theory~\cite{cover,ash} can fit in the argument based on operations naturally, since the amount of
information, such as the Shannon entropy~\cite{shannon}, is defined through probabilities.

Moreover, information processing can also be seen as a physical operation, since once information
is encoded in a physical state any computational manipulation is realized as an operation on the
state~\cite{turing,turing_correction}. This way, we can construct an operational scenario,
incorporating the notion of probability via information with thermodynamics~\cite{HMS}.

As a concrete example, here we consider the derivation of the Boltzmann distribution in the
canonical ensemble. Perhaps its most notable derivation using the concept of information (or
entropy) is the one by Jaynes~\cite{Jaynes}, who claimed the principle of maximum entropy (PME). Jaynes
identified the equilibrium as the state that maximizes the Shannon entropy with respect to the
probability of each microscopic configuration under the constraint on the total energy.

While Jaynes' approach has been very successful, the PME is essentially based on the principle of
equal \textit{a priori} probabilities (Bayesian view of probability). This means that no
operations are involved in the \textit{a priori} probabilities for the premise of the PME, unlike
in those of frequentism.

More recent work that may be relevant is the formulation of the canonical ensemble in the language
of quantum mechanics \cite{goldstein2006,popescu2006}. They showed that the state $\rho$ of a
small system is approximately equal to the canonical state $\exp(-H/k_B T)$, as a result of
entanglement between the system and its environment, provided the interaction between the system
and the environment is weak. Here, $H$, $k_B$, and $T$ are the Hamiltonian of the system, the
Boltzmann constant, and the temperature of the environment. Their results are very smart and
elegant in their own right, however, they have assumed the \textit{a priori} equiprobability and
it is still unclear whether the consideration of quantum entanglement is requisite for the
foundations of statistical mechanics.

In this paper, we derive the Boltzmann distribution for the canonical ensemble in an operational
manner, i.e., constructing an operation-based scenario, with which we define a function to discuss
equilibrium. This approach is useful to clarify the role of information, albeit implicit, in what
we already see as a common sense in physics.

A key ingredient in our work that brings the notion of information into physics is information
processing, or more specifically, information erasure. The physics of information erasure
clarified the link between thermodynamic and information-theoretic entropies
\cite{landauer,bennett82,bennett_rev,zurek1989,shizume1995,maruyama09,mandal2012}, and it played a
central role in resolving the paradox of Maxwell's demon. It states that the erasure of one bit of
information (in the demon's memory) requires a work consumption of at least $k_B T\ln2$. Here,
$k_B$ is the Boltzmann constant and $T$ is the temperature of the heat bath with which the memory
system is in contact. Incidentally, despite the extremely small value of $k_B\ln2$, which is
roughly $1\times 10^{-24}$ J/K, strong experimental evidence for the information erasure principle
has recently been reported \cite{berut2012,jun2014,roldan2014,koski2014}. If the information
content in an $N$-bit string is $NH(p)<N$, where $H(p)=-p\log_2 p -(1-p)\log_2 (1-p)$ is the
Shannon entropy, then the minimum work for erasure becomes $Nk_B TH(p)\ln2$, as shown in Ref.
\cite{HMS}. This is because the optimal data compression makes the length of the string from $N$
to $NH(p)$, and after this compression we erase information in the $NH(p)$ bits in which 0 and 1
appear with equal probability, spending $Nk_B TH(p)\ln2$ of work.

Here, we make the demon play as a symbolic entity that carries out operations, as we shall present
below. Also, because the definition of equilibrium is independent of operations, our scenario has
a potential to be applied to nonequilibrium statistical mechanics, as we will describe briefly,
taking the fluctuation-dissipation theorem \cite{kubo} as an example.

%%%%%%%%%%%%%%%%%%%%%%%%%%%%%%%%%%
\section*{Result}
%%%%%%%%%%%%%%%%%%%%%%%%%%%%%%%%

Let us clarify first what we mean by ``Maxwell's demon", as sometimes this can be a source of
confusion. We basically follow the original idea by Maxwell \cite{maxwell1867}, although our demon
does not intend to violate the second law of thermodynamics~\cite{demon2,norton1998,norton1999,bub2001,Parrondo2015}.

In this paper, the demon is an entity that can measure and change the energy levels of particles,
and manipulate/process information encoded in memory registers (cells). As it will be clearer
below, the particles can have only two distinct energy levels and this is what the demon measures
and handles. The demon can of course access the heat bath, thus extract and discard energy from/to
it via appropriate tools, complying with the laws of thermodynamics.

The memory is embedded on a long tape, as in the Turing machine that is an abstract, but common,
model for information processing. The tape can also be used as a working space for computation, if
necessary.

We note that the demon should be able to work autonomously, once the protocol and algorithm for
its task are given. The phrase ``autonomous system" may refer to a system consisting of mechanical
parts that is designed to work on its own (without energy supply or active control from outside),
e.g., a Szilard-engine-type machine presented in Ref\cite{lu2014}. Nevertheless, for our purpose,
it is sufficient to consider a system that proceeds deterministically reacting to the input from
outside, complying with physical laws. Naturally, in order to work independently, it should not be
fed any extra information or energy as a whole.

So, the name ``demon" has merely a symbolic meaning here; it can be replaced with a machine that
is capable of storing and processing information, and manipulating the particle states. Although
it could be done with some inspiration from an example in Ref\cite{lu2014}, devising such a
structure in detail is out of our scope and would be left for future work.

%%%%%%%%%%%%%%%%%%%%%%%%%%%%%%%%%%
\subsection*{Thermo-Turing model}
%%%%%%%%%%%%%%%%%%%%%%%%%%%%%%%%
The primary components of our model are a set $\phys$ of $N$ particles with two energy levels
$E_0$ and $E_1 (>E_0)$ and a long tape $\tape$, which represents the demon's memory and contains a
sequence of $N$ memory cells. We let $\Phi_0, \Phi_1$ and $\epsilon$ denote the two (ground and
excited) states of the particles and the energy gap $E_1-E_0$, and assume that each particle is
numbered to make a correspondence with a memory cell. The memory tape $\tape$ can be thought of as
a part of the demon and it is very similar to the one we typically consider in the context of
Turing machine. Each memory cell can store a binary information, either 0 or 1, and it can be
modelled as the Szilard engine \cite{szilard1929}, which is a one-molecule gas with a partition at
the center of cylinder. We call the mechanism comprising of $\tape$ and the demon a
``thermo-Turing model" in the following discussion.

In the context of (thermodynamic and algorithmic) entropy from the operational point of view,
Zurek considered a model of demon with a Turing machine in Ref. \cite{zurek1989}. Here, we
incorporate the notions of information processing a la Turing and of thermodynamic consideration
of Maxwell's demon to step in the field of statistical mechanics.

\begin{figure}[t]
\begin{center}
\includegraphics[scale=1.4]{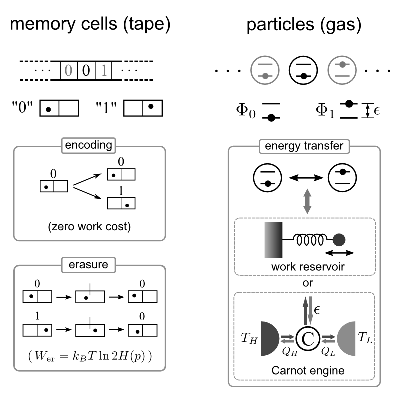}
\caption{{\bf Elements of the thermo-Turing model.} Two basic operations on the memory cells,
encoding and erasure, are depicted on the left. Each memory cell is modeled with a single molecule
gas contained in a cylinder. Encoding is carried out by slowly displacing the region of volume
$V/2$ in which the molecule is kept, thus no work is consumed, or by simply rotating the cylinder
by $180$ degrees when the value is one. Information erasure requires a work consumption of $k_B
T\ln2 H(p)$, where $H(p)$ is the amount of information stored in the tape. The two-level particle
system is sketched on the right. Energy transfer to the particles can be done with a work
reservoir and an unspecified mechanism, or with a fictitious Carnot engine. All these operations
on the memory cells and the particles are controlled by the demon (not shown).}
\label{fig:cycle_elements}
\end{center}
\end{figure}

In our thought experiment, the interaction between $\phys$ and the heat bath is mediated by the
demon (or the thermo-Turing machine). The rough idea is as follows. The interaction with heat bath
causes noise on $\phys$ and an energy change in it. The degree of noise depends on the bath
temperature $T$, but we represent it only by probability $p$ of a state flip. Equilibrium is
defined as the state in which the energy change in $\phys$ is balanced with the energy consumption
for subsequent manipulations of the memory tape $\tape$ at $T$. Thus, the temperature $T$ comes in
to the discussion explicitly only through demon's actions on $\phys$. Note that it is legitimate
to assume that $\tape$ is designed to make the stored information insensitive to thermal
fluctuation. This picture (of having a direct effect of $T$ on $\tape$) may appear strange from
the viewpoint of the conventional deductive approach. However, this scenario allows us to use the
demon as a subject of physical 'operations' to bring thermodynamic notions into the discussion.
The elements of the thermo-Turing model and basic operations therein are depicted in Fig.
\ref{fig:cycle_elements}.

Naturally, we consider the memory tape $\tape$ to let it reflect the state of particles in
$\phys$. Suppose a situation in which the fraction $p$ of a set of $N$ particles are in the
excited state, i.e., $pN$ particles are in $\Phi_1$, while $(1-p)N$ in $\Phi_0$. Let $F$ be the
amount of work that the entire system ($\phys + \tape$) can potentially exert towards the outside,
when we let it be in the state where all particles are in $\Phi_0$ and all memory cells store `0'.
It simply means that we take the state with all in $\Phi_0$ and `0' as the origin for the quantity
$F$.

The energy stored in $\phys$ contributes to $F$ positively, and its amount is $E:=pN\epsilon$. On
the other hand, in order to erase all information on the tape, we need to consume some energy
$W_\mathrm{er}$. As a result, we have
\begin{equation} \label{f_defin}
    F=E-W_{\mathrm{er}}.
\end{equation}
Since we are naturally interested in the optimal (largest) value of $F$ for a given $p$ in order
to characterize the state uniquely, $W_\mathrm{er}$ needs to be minimized. Thus, we have
$W_\mathrm{er} =N H(p) k_{B} T \ln 2$ \cite{HMS}, which leads to
\begin{equation}\label{the_f_function}
    F = p N \epsilon - NH(p) k_{B} T \ln 2.
\end{equation}
Equation (\ref{the_f_function}) resembles the Helmholtz free energy, i.e., $F=U-TS$, however, the
conceptual difference behind them should be emphasized. The point is in presenting the operational
scenario for statistical mechanics by identifying the thermodynamic entropy with the information
entropy.

With the definition of $F$, which is computable for any physical state, we shall now define the
equilibrium in terms of $F$. We call that the state is in equilibrium when its $F$ is stationary,
i.e.,
\begin{equation}\label{equilibriumcondition}
\Delta F=0
\end{equation}
against small noises on the particles. We consider the NOT (flipping) operation on a particle as
an elementary process of the noise, thus Eq. (\ref{equilibriumcondition}) is a condition against a
small number of random NOT operations on $\phys$. This definition of equilibrium is associated
with the stationarity of the principal system and the memory tape, rather than the largest
likeliness of the state as in Jaynes' argument \cite{Jaynes}. Our definition fits the operational
point of view better, because the quantity $F$ can be computed by considering physical operations.
The operational process (by the demon) will be presented, after deriving the expression of the
Boltzmann distribution. Also, a comparison with Jaynes' work is given in Supplementary Material.

Let us compute $\Delta F$ for a probability change from $p$ to $p^\prime$.
\begin{equation}\label{DeltaFagainstDp}
\Delta F = F^\prime - F = \Delta p \cdot N \epsilon - N k_B T \ln 2\cdot\Delta H(p),
\end{equation}
where $\Delta p = p^\prime - p$ and $\Delta H(p) = H(p^\prime) - H(p)$.  Since the number of
errors is small ($\Delta p\ll 1$), $\Delta H(p)=dH(p)/dp \Delta p \equiv H^{\prime} (p) \Delta p$.
The equilibrium condition, $\Delta F=0$, gives
\begin{equation}\label{balance}
    \epsilon - k_B T \ln 2 \cdot H^{\prime} (p) = 0.
\end{equation}

Suppose that the probability change, $p\rightarrow p^\prime$, is induced by thermal noise that
flips the state of a randomly chosen particle in $\phys$ between $\Phi_0$ and $\Phi_1$. Noting
that $\ln 2 \cdot H^{\prime} (p) =\ln [ (1-p)/p ]$, we see that Eq. (\ref{balance}) reduces to
\begin{equation}
    \frac{p}{1-p}=\exp \left( -\frac{\epsilon}{k_{B}T} \right),
\label{Boltzmann}
\end{equation}
which is nothing but the Boltzmann distribution. The generalization of the model to $d$-level
physical systems is presented later in this section.

A similar analysis on the effect of the Toffoli gate is also insightful, however, it is summarized
in the Supplementary Material so that we focus on the derivation of the Boltzmann distribution
here.

\begin{figure}[t]
\begin{center}
\includegraphics[scale=1]{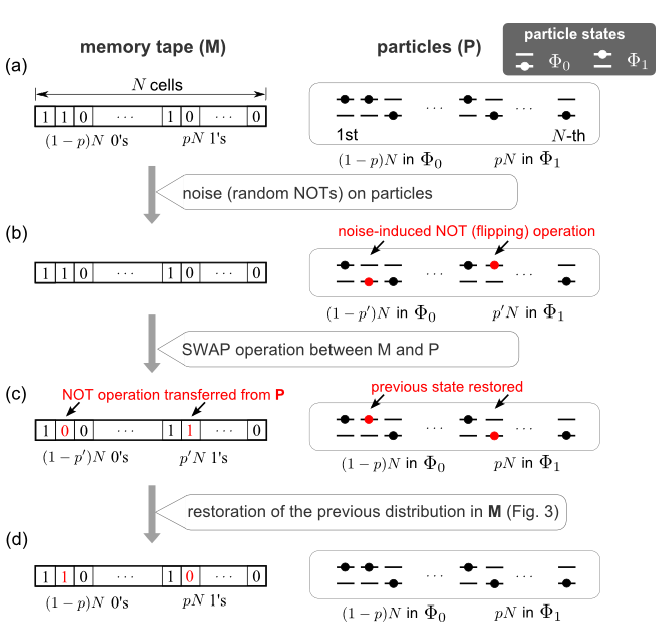}
\caption{{\bf The virtual process for which we consider the change of $F$.} (a) Each memory cell
has a perfect correlation with the corresponding particle's state. (b) The interaction between the
particles and heat bath causes a NOT (=flipping) operation to a small number of particles
randomly. (c) The demon swaps the information stored in the memory and the particle state, e.g.,
0-$\Phi_1$ becomes 1-$\Phi_0$, and vice versa. (d) Both $\tape$ and $\phys$ return to the original
state that is the same as (a).} \label{fig:cycle}
\end{center}
\end{figure}

Next, we describe the operational scenario that naturally leads to the equilibrium condition,
$\Delta F=0$ with Eq. (\ref{DeltaFagainstDp}). Figure \ref{fig:cycle} depicts the process.

\begin{figure}[t]
\begin{center}
\includegraphics[scale=0.9]{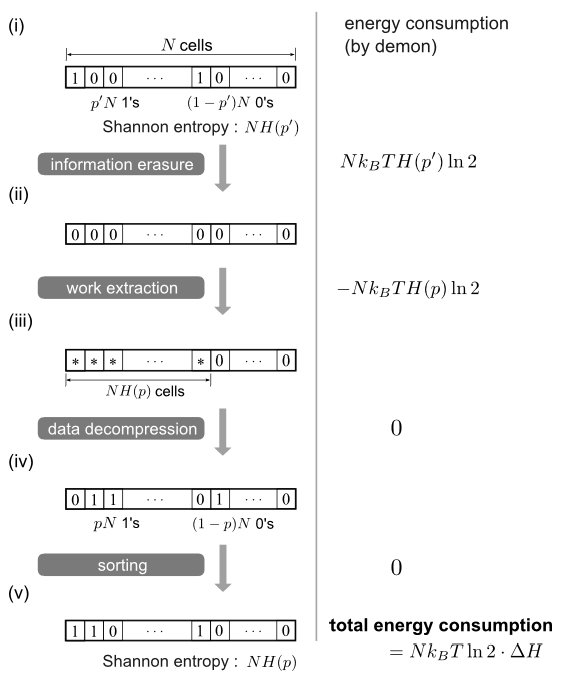}
\caption{{\bf State transformation to change the probability distribution from $p^\prime$ to $p$.}
All memory cells are reset to `0', consuming $k_BT\ln2 H(p^\prime)$ of work (from (i) to (ii)). In
(iii), the one-molecule gases in $NH(p)$ cells are expanded isothermally, giving the demon
$k_BT\ln2 H(p)$ of work. The `$*$' sign in (iii) represents a randomized memory state with no
physical distinction between $0$ and $1$; the molecule can move around in the whole configuration
space of the cylinder. Since the state in (iii) is the same as the resulting state of data
compression for an $N$-bit string containing $NH(p)$ bits of information, data decompression leads
to the string in which $pN$ bits are in `1' and the rest are in `0' as in (iv). By permutating the
bit string, which can be done for free of energy, the memory tape with $NH(p)$ of information and
the perfect correlation with the particles' state can be realized.} \label{fig:transform}
\end{center}
\end{figure}

\begin{enumerate}

\item[(a)] One-to-one correspondence between the data stored in the memory tape $\tape$ and the
state of each particle in $\phys$ is established. That is, if a particle is in the ground state
$\Phi_0$ the corresponding memory cell stores `0', and if the particle is in $\Phi_1$ the memory
has `1'. This correspondence can be made by measuring the particle state and copying the result to
the memory, which can be done without energy consumption \cite{bennett82,bennett_rev}.

\item[(b)] During some time interval $\Delta t$, a NOT (flipping) operation is applied to a few
randomly chosen particles. This may be induced by noise or thermal fluctuation, i.e., the
interaction between particles and the heat bath. Since the interaction with heat bath is not under
the demon's control, he spends zero energy here. $\Delta t$ can be taken so that the number of
flips is much smaller than $N$.

\item[(c)] The demon swaps the states of particles and memory registers in the tape. The effect of
the NOT operations in (b) is now transferred to the tape, while the state of particles is restored
to be the one in (a). The energy of $\Delta p N\epsilon=(p^\prime-p)N\epsilon$ is acquired by the
demon to change the particle state, while the SWAP operation for information can be performed
without energy consumption as it keeps the entire entropy unchanged. $\tape$ now has the Shannon
entropy $H(p^\prime)$.

\item[(d)] The demon transforms the Shannon entropy of $\tape$ from $H(p^\prime)$ to $H(p)$. This
can be done by the process depicted in Fig. \ref{fig:transform}, which is explained in detail
below. The energy required for this entropy transformation is $N k_B T\ln2\cdot \Delta H:=N k_B
T\ln2 (H(p^\prime)-H(p))$.

\end{enumerate}


The resulting state in (d) of the above process is the same as (a), and all steps can be made
completely autonomous. That is, no traces of the actions by demon are left not only inside $\tape$
and $\phys$, but also in their surrounding environment, while the only possibility of the trace is
the amount of energy the environment received. Therefore, for the the joint thermo-Turing system
$\tape + \phys$ to be in equilibrium, i.e., no macroscopic change detectable from outside, the
energy transfer between the joint system and the environment should be zero. Indeed, this
condition can be written as
\begin{equation}
\Delta p N\epsilon - N k_B T\ln2\cdot \Delta H=0,
\end{equation}
which is $\Delta F=0$.

Figure \ref{fig:transform} shows the process to change the state of memory tape so that its
entropy is transformed from $H(p^\prime)$ to $H(p)$. Incidentally, this process can be seen as a
special (classical) case of the one in Fig. 2 of Ref. \cite{maruyama05}, which presented a
thermodynamical transformation of quantum state from $\sigma$ to $\rho$. Here, let us proceed with
Fig. \ref{fig:transform} solely.

From Fig. \ref{fig:transform} (i) to (ii), the demon erases all the information stored in the
tape, consuming at least $Nk_B T\ln2 \cdot H(p^\prime)$ of work.

In Fig. \ref{fig:transform} (iii), the demon extracts $Nk_B T\ln2 \cdot H(p)$ of work from the heat bath
by letting the gas in $NH(p)$ cells expand isothermally, which is possible since each memory cell
can be modelled by a one-molecule gas. Note that the demon can always have the values of $p$ and
$p^\prime$ since measurement can be done for free. Inserting a partition at center of each cell,
now there are $NH(p)/2$ cells that represent `0' and another $NH(p)/2$ cells `1' (with negligible
fluctuation when $N$ is large enough).

Then the (Shannon) data decompression \cite{shannon} is performed on all the $N$ cells to have
$pN$ cells in `1' and $(1-p)N$ cells in `0' as in Fig. \ref{fig:transform} (iv). Now that the
number of cells in `1' is the same as that of particles in excited state $\Phi_1$ in Fig.
\ref{fig:cycle} (d), the demon can sort the order of memory cells to make one-to-one
correspondence with the particle states. The sorting process can be done isentropically, thus
autonomously, since it is achieved simply by applying an appropriate permutation. Alternatively, a
controlled-NOT operation may be applied between $\tape$ and $\phys$, with a memory cell as a
control bit and the corresponding particle as a target bit. Because the number of particles in
$\Phi_1$ is the same at Steps (iv) and (v), no extra energy is necessary as a whole.

In summary, we have devised a physical scenario with which we can derive the Boltzmann equilibrium
distribution in the statistical mechanics in an operational manner. The operations are performed
on the particles and a virtual Turing-machine-type memory cells. We have symbolically used
Maxwell's demon-type intelligent being as the principal operator, but all actions are autonomous
and leave no traces observable from outside, thus the demonic actions can be programmed in the
Turing machine per se.

The erasure principle, stemming from the paradox of Maxwell's demon, bridges thermodynamics and
statistical mechanics via the notion of probability in information theory. It should be emphasized
that we did not base our argument on the equiprobability principle. That is, we did not rely on
the standard micro-canonical statistical mechanics, in which the entropy $S$ is given by the
Boltzmann formula $S=k_B\log \Omega(E)$ with $\Omega(E)$ being the number of states under a given
energy $E$.

Also, the above model can be used to justify the equivalence between thermodynamic and information
theoretic entropies, which was discussed in our previous work\cite{HMS} in a different context. A
brief argument is given in this line in Supplementary Material.

%%%%%%%%%%%%%%%%%%%%%%%%%%%%%%%%%%
\subsection*{Generalization to $d$-level system}
%%%%%%%%%%%%%%%%%%%%%%%%%%%%%%%%%%
The above argument to derive the Boltzmann distribution can be generalized to the systems of
arbitrary levels. That is, the cells of the tape can store $d$ values from 0 to $d-1$, and there
are $d$ possible states for particles, $\Phi_0, \Phi_1,...,\Phi_{d-1}$, whose energy levels are
$E_0, E_1,..., E_{d-1}$, respectively. Let $p_i$ be the ratio of the number of particles in the
state $\Phi_i$. Suppose that the $k$-th cell of the tape stores the value $i$ when the $k$-th
particle is excited to $\Phi_i$. This state preparation can be completed by simply copying the
measurement result about the particle state. %We will use another notation $\carnot_i^{-1}$ to
%indicate the engine $\carnot_i$ operated in the reverse direction.

Instead of the noise-induced random NOT studied above, let us consider random SWAP operations that
change the state of a particle. Let $\mathrm{SWAP}_{ij}$ denote a SWAP between two states $\Phi_i$
and $\Phi_j$, namely, $\mathrm{SWAP}_{ij}$ maps the state $\Phi_i$ of a randomly chosen particle
to $\Phi_j$ and vice versa. Note that the NOT operation between two levels is effectively the same
as the SWAP between them, so the process for the thermo-Turing system is basically the same as the
one described above (and Fig. \ref{fig:cycle}) with the replacement of NOT with SWAP.

Suppose that a $\mathrm{SWAP}_{ij}$ has occurred to one of the particles. The $\mathrm{SWAP}_{ij}$
changes its state if it is in either $\Phi_i$ or $\Phi_j$, otherwise nothing happens. The
probability of such a `successful' $\mathrm{SWAP}_{ij}$ is $p_i+p_j$. After the demon swaps the
information between $\tape$ and $\phys$, the memory tape $\tape$ contains information after the
$\mathrm{SWAP}_{ij}$, and the particles $\phys$ returns to the state before the
$\mathrm{SWAP}_{ij}$ (as in Step (c) above). Thus energy change in $\phys$ due to this operation
is $(E_j-E_i)(p_i-p_j)/(p_i+p_j)$ on average.

The change in the erasure entropy times temperature after the single $\mathrm{SWAP}_{ij}$ and swap
between $\tape$ and $\phys$ is
\begin{align}
&Nk_{B}T\ln2 \left[\frac{p_i}{p_i+p_j}H\left(p_i-\frac{1}{N},p_j+\frac{1}{N}\right)\right. \notag \\
&\left.+\frac{p_j}{p_i+p_j}H\left(p_i+\frac{1}{N},p_j-\frac{1}{N}\right)\right]-Nk_{B}T \ln2 \cdot H(p) \notag \\
& \ \ \ = k_{B}T\ln2
\frac{p_i-p_j}{p_i+p_j}\ln\frac{p_i}{p_j}+\mathcal{O}\left(\frac{1}{N}\right),
\end{align}
where
\begin{align}
& H\left(p_i\pm\frac{1}{N},p_j\mp\frac{1}{N}\right) =  \nonumber \\
& -\sum_{k\neq i,j}p_k\log_2 p_k - (p_i\pm\frac{1}{N})\log_2 (p_i\pm\frac{1}{N})
-(p_j\mp\frac{1}{N})\log_2 (p_j\mp\frac{1}{N}).
\end{align}

Making the change in $F$ equal to zero as in the case of bits and two-level particles, we arrive
at the desired relation:
\begin{align}
(E_j-E_i)\frac{p_i-p_j}{p_i+p_j}-k_{B}T\frac{p_i-p_j}{p_i+p_j}\ln \frac{p_i}{p_j} & = 0 \notag \\
\frac{p_j}{p_i} & = \exp\left(-\frac{E_j-E_i}{k_{B}T}\right). \label{general_eq}
\end{align}
This relation holds for any pairs of $i$ and $j$, hence $p_i\propto \exp (-E_i/k_B T)$ for all
$i$.

%%%%%%%%%%%%%%%%%%%%%%%%%%%%%%%%%%
\section*{Discussion}
%%%%%%%%%%%%%%%%%%%%%%%%%%%%%%%%

In principle, our thermo-Turing model can be applied to generic non-equilibrium processes, as far
as we can assume that the operations by demon can be carried out sufficiently fast, compared with
the dynamics. Here, we present a modest step to this direction, choosing a particular model which
exhibits a characteristic feature of fluctuation-dissipation theorem.

Suppose that a spatially fluctuating external field that works as a perturbation to energy levels
is applied to let the system $\phys$ deviate from macroscopic equilibrium. This field causes a
small change to the energy gap of the particle at the $n$-th site to make it $\epsilon - u_n$, and
we assume $u_n\ll \epsilon$ and $\sum_n u_n=0$ for simplicity. Such a change may be seen as a
result of the Stark or the Zeeman effect, but we do not need to specify the origin of the shift
for our discussion.

In order to discuss statistical quantities for each particle, the site $n=1, 2,\dots, N$ should be
regarded as a block, which consists of sufficiently many members. For the $n$-th block, due to the
energy shift $u_n$, the local equilibrium distribution becomes
\begin{equation}\label{pleq}
    p_\mathrm{leq}(u_n)=\left[1+\exp \left(\frac{\epsilon-u_n}{k_{B}T} \right)\right]^{-1}.
\end{equation}
Under this distribution, the operations by demon within the $n$-th block balance with the external
field. The index `leq' for $p$ in Eq. (\ref{pleq}) stands for local equilibrium.

What we are interested in is the amount of dissipation, given the fluctuation of $\{u_n\}$,
$\mathcal{F}=\sum_n u_n^2$. Imagine that the demon now looks at all the blocks as a whole, and
attempts to make all particles return to the same equilibrium state, i.e., $u_n=0$ for all blocks.
This is done by changing the energy state of each block and erasing information about the spatial
variation of the energy shifts. The work that needs to be done by demon in order to change the
distribution $p_\mathrm{leq}$ to that of equilibrium
$p_\mathrm{eq}=\left.p_\mathrm{leq}\right|_{u_n=0}$ is
\begin{eqnarray}\label{averagework}
\ave{W} &=& \sum_n \left[p_\mathrm{eq}\epsilon - p_\mathrm{leq}(u_n)(\epsilon - u_n) \right] - k_B
T\ln2 \sum_n \left[H(p_\mathrm{eq})-H(p_\mathrm{leq})\right] \nonumber
\\
&=& \sum_n \left[p_\mathrm{eq}\epsilon - \left(p_\mathrm{eq}+p_\mathrm{leq}^\prime(0) u_n +
\frac{1}{2}p_\mathrm{leq}^{\prime\prime}(0)u_n^2 +\mathcal{O}(u_n^3)\right)(\epsilon-u_n)\right] \nonumber \\
& &
- k_B T\ln2 \sum_n \left[H(p_\mathrm{eq}) - H\left(p_\mathrm{eq}+p_\mathrm{leq}^\prime(0) u_n + \frac{1}{2}p_\mathrm{leq}^{\prime\prime}(0)u_n^2 +\mathcal{O}(u_n^3)\right)\right] \nonumber \\
&=& \sum_n \left[ -p_\mathrm{leq}^\prime(0)\epsilon u_n + p_\mathrm{leq}^\prime(0) u_n^2 +
p_\mathrm{eq}u_n -\frac{1}{2}p_\mathrm{leq}^{\prime\prime}(0)\epsilon u_n^2 \right] \nonumber \\
& & + k_B T\ln2 \frac{dH}{dp} p_\mathrm{leq}^\prime(0)\sum_n u_n + \frac{1}{2} k_B T\ln2 \left(
\frac{d^2 H}{dp^2} (p_\mathrm{leq}^\prime(0))^2
+ \frac{dH}{dp} p_\mathrm{leq}^{\prime\prime}(0)\right) \sum_n u_n^2 +\mathcal{O}(\sum_n u_n^3)\nonumber \\
&\simeq& \left[-\frac{1}{2}\left(\epsilon - k_B T\ln2
\frac{dH}{dp}\right)p_\mathrm{leq}^{\prime\prime}(0) + \left(p_\mathrm{leq}^\prime(0) +
\frac{1}{2}k_B T \ln2 \frac{d^2 H}{dp^2}
(p_\mathrm{leq}^\prime(0))^2\right)\right]\cdot \sum_n u_n^2 \nonumber \\
&=& p_\mathrm{leq}^\prime(0)\left(1+\frac{1}{2}k_B
T\left(-\frac{1}{p(1-p)}\right)p_\mathrm{leq}^\prime(0)\right)\cdot\sum_n u_n^2 \nonumber \\
&=& \frac{1}{2} p_\mathrm{leq}^\prime(0) \sum_n u_n^2,
\end{eqnarray}
where $p_\mathrm{leq}^\prime(0):=\left.\frac{dp_\mathrm{leq}}{du_n}\right|_{u_n=0}$,
$p_\mathrm{leq}^{\prime\prime}(0):=\left.\frac{d^2p_\mathrm{leq}}{du_n^2}\right|_{u_n=0}$, and
$\frac{dH}{dp}$ and $\frac{d^2H}{dp^2}$ are evaluated at $p=p_\mathrm{eq}$. Also, we have used
$\sum_n u_n=0$ in the fourth equality, and the condition Eq. (\ref{balance}) for equilibrium in
the fifth equality. From Eq. (\ref{pleq}), we have
\[
p_\mathrm{leq}^\prime(0)=\frac{1}{k_BT}\frac{\exp{(\epsilon/k_B T)}}{(1+\exp{(\epsilon/k_B
T)})^2}>0,
\]
therefore,
\begin{equation}\label{FD}
\langle W \rangle =\frac{1}{2}p_\mathrm{leq}^\prime(0) \mathcal{F} >0.
\end{equation}
Equation (\ref{FD}) means that the response of the system to the external field results in the
positive work by demon, $\ave{W}>0$, which is dissipated into the heat bath. Further, it is
proportional to the fluctuation of the external field. It is a simple expression of the
dissipation-fluctuation theorem in the linear approximation of the fluctuating potential $u_{n}$.
Also, Eq. (\ref{FD}) is interesting in the sense that our model explicitly takes into account of
the cost of `forgetting the past', which is simply neglected in the standard consideration of
Markovian processes.

The readers who are familiar with the standard dissipation-fluctuation theorem \cite{kubo} would
feel more comfortable with the fluctuation in time rather than the spatial one of the external
potential. In that case, one can reorder the site numbers according to the order of occurrence of
$u_n$. Then, the $n$ can be interpreted as time, and the average $\langle \cdot \rangle$ can be
understood as that over a long time.

%%%%%%%%%%%%%%%%%%%%%%%%%%%%%%%%

\section*{Acknowledgments}
The authors acknowledge the Yukawa Institute for Theoretical Physics at Kyoto University on the
YITP workshop YITP-W-11-25 on ``Hierarchy in Physics through Information -- Its Control and
Emergence --". K.M. acknowledges support from the JSPS Kakenhi (C) Grants No. 26400400. This work
was partially supported by the NINS Youth Collaborative Project and DAIKO Foundation.

%\section*{Author Information}
%\subsection*{Author contributions}
%A. H. provided the preliminary idea and results on the thermo-Turing model. K. M. and Y. S.
%clarified the idea and results. Further, K. M. provided the generalization of the thermo-Turing
%model. All authors contributed to writing the manuscript.
%
%\subsection*{Competing financial interests}
%The authors declare no competing financial interests.

\clearpage

%====================================================================================================%
%====================================================================================================%
%====================================================================================================%

\renewcommand{\theequation}{S\arabic{equation}}
\setcounter{equation}{0}

%\begin{titlepage}
\begin{center}
\large{\bfseries Supplementary Material: Operational derivation of Boltzmann distribution with
Maxwell's demon model}
\end{center}%\vspace{20pt}

%\end{titlepage}

%%%%%%%%%%%%%%%%%%%%%%%%%%%%%%%%%%
\section{Comparison with Jaynes' Work \label{ack02}}
%%%%%%%%%%%%%%%%%%%%%%%%%%%%%%%%%%
In the work on the relation between information theory and statistical mechanics, Jaynes derived
the corresponding equilibrium state by postulating the principle of maximum entropy~\cite{Jaynes},
which is described as follows. The plausible probability distribution is determined by the
requirement that the estimator function ${\rm Est} (p_1, p_2, \dots, p_n)$ is maximized under a
given average energy $\ave{E} = \sum_i p_i E_i$ with $\sum_i p_i = 1$. Following
Shannon~\cite{shannon}, he required the mathematical properties for the estimator ${\rm Est} (p_1,
p_2, \dots, p_n)$: (i) ${\rm Est} (p_1, p_2, \dots, p_n)$ is a continuous function of $p_i$, (ii)
$f(n) := {\rm Est} (1/n, 1/n, \dots, 1/n)$ is a monotonic increasing function of $n$, and (iii)
${\rm Est} (p_1, p_2, \dots, p_n)$ satisfies the composition law. From these mathematical
requirements (i)-(iii), the estimator is uniquely determined to be equal to the Shannon entropy,
i.e., ${\rm Est} (p_1, p_2, \dots, p_n) = - \sum_i p_i \log p_i$. Maximizing the Shannon entropy
under the constraints that $\ave{E}$ is fixed and $\sum_i p_i = 1$, the plausible probability
distribution is obtained to be the one in Eq. (19) of the main text. Note that the temperature
does not come in from the physical assumption but is defined so that the free energy thereby
obtained coincides with the Helmholtz free energy.

In the present work, we have devised a model consisting of a set of particles (physical system)
and a demon with components for information processing, namely a memory tape and a processor. The
memory tape is in the contact with a heat bath, and the demon operates them to simulate the effect
of interaction between the particle system and a heat bath, so that the particles' energy state
will have the right probability distribution. We have defined the equilibrium state operationally
and subsequently derived the Boltzmann distribution. Thus, the temperature dependence of the
distribution was naturally obtained, whereas it was not so in Jaynes' mathematical work.

\section{Effect of Toffoli gates}
The effect of the Toffoli gate as an error operation is also worth studying, since errors can be
any arbitrary logical operations and it is well known in computer science that arbitrary logical
operations can be simulated by NOT and Toffoli gates as far as the operation is logically
reversible \cite{nielsenchuang}. The Toffoli gate works on three bits, two of which are control
bits and the other one is the target bit. The state of the target particle is flipped when the two
control bits are in the excited state (`1' in the tape $\tape$), thus the flip of the target is
activated with probability $p p^*$, where $p^*=(pN-1)/(N-1)$ is the probability of finding the
second control particle in the excited state $\Phi_1$ when the first one is also in $\Phi_1$. The
flip operation on the target is applied to the particle in $\Phi_1$ with probability
$p^{**}=(pN-2)/(N-2)$ and the resulting ratio of those in $\Phi_1$ will be $p-1/N$. Similarly, if
the target was in $\Phi_0$ the ratio $p$ will be changed to $p+1/N$. Therefore, the average change
of the Shannon entropy is
\begin{eqnarray}\label{DeltaH2ndOrder}
\Delta H &=&
pp^*\left[p^{**}H\left(p-\frac{1}{N}\right)+(1-p^{**})H\left(p+\frac{1}{N}\right)\right] +
(1-pp^*)H(p)-H(p) \nonumber \\
&\simeq& \frac{1}{N}pp^* (1-2p^{**})H^\prime(p) + \frac{pp^*}{2N^2} H^{\prime\prime}(p),
\end{eqnarray}
where the second term $(1-pp^*)H(p)$ represents the case of no activation of the flip on target.

With the above $\Delta H$, the change in $F$ due to a random Toffoli operation can be computed as
\begin{eqnarray}
\Delta F ( {\rm Toffoli} ) &=& \Delta p N\epsilon - k_B T\ln2\cdot \Delta H \nonumber \\
&=&  pp^*(1-2p^{**}) \epsilon -pp^*(1-2p^{**}) k_B T\ln2\cdot H^\prime(p) - \frac{pp^*}{2N}
k_B T \ln2 \cdot H^{\prime\prime}(p) \nonumber \\
&=& - \frac{pp^*}{2N} k_B T \ln2 \cdot H^{\prime \prime} (p) + \mathcal{O} \left( \frac{1}{N^2}
\right)
\nonumber \\
&\rightarrow&  - \frac{p^2}{2N} k_B T \ln2 \cdot H^{\prime \prime} (p) \;\;\;(\mbox{as}\; N\gg 1).
\label{Toffoli_def}
\end{eqnarray}
We have used the equilibrium condition, Eq. (5) in the main text, in the third equality. The
effect of the random Toffoli gate is essentially the same as that of the random NOT (corresponding
to $\Delta H$ with $pp^*=1$ in Eq. (\ref{DeltaH2ndOrder})) except for the overall probability
factor $p^2$ for having two $1$'s in the control bits.

Thus, the sum of the second order contribution from the simple random NOT (=flip) and Toffoli
operations is
\begin{equation}
\Delta F = -k_B T \frac{\ln 2}{2N}\left[ \# ( {\rm NOT} ) + \# ( {\rm Toffoli} ) p^2\right]
H^{\prime \prime} (p)  \;(> 0). \label{Toffoli}
\end{equation}
That is, the change $\Delta F$ by the random NOT and Toffoli gates strengthens the local stability
of the cost function $F$ near the equilibrium, which is also expected from the convexity of the
entropy function. This further supports our definition of equilibrium in the thermo-Turing model.
It is also worth noting that the quantity in the square bracket is the effective computational
complexity \cite{zurek1989}.

\section{Note on thermodynamic and information-theoretic entropies}
%%%%%%%%%%%%%%%%%%%%%%%%%%%%%%%%%%
In our previous paper~\cite{HMS}, using the scenario of Maxwell's demon, we showed that the
thermodynamic entropy coincides with the information theoretic entropy in the optimal case of the
memory reset. In this section, we attempt to look at the same problem from a different perspective
going back to the the basic thermodynamics.

Let us recall how the thermodynamic entropy was introduced in the context of Carnot's theorem. Let
the work exerted towards the outside be $W(i\rightarrow f)$ for an isothermal state change
$i\rightarrow f$, which is in general different from the change of the internal energy
$U_{i}-U_{f}$. An invisible energy flow that contributes to the energy balance is called heat $Q$
exchanged between the system and the heat bath. Thus, the energy conservation is written as
\begin{equation}
W(i\rightarrow f) = U_{i}-U_{f}+Q. \label{work}
\end{equation}
Carnot's theorem claims that the maximal heat flow from the heat bath in isothermal process is
proportional to the temperature $T$ of the heat bath,
 \begin{equation}
Q_{max} = T S_{th},
 \label{entropy}
\end{equation}
where the maximization is made over all possible intermediate processes between the initial and
the final states~\cite{TASAKI}. The coefficient $S_{th}$ is defined as the increase of the
thermodynamic entropy.

Now go back to our thermo-Turing model. Clearly, our $F$ in Eq. (2) of the main text is the
negative of the work that can be exerted outwards,
\begin{equation}
F=-W(i\rightarrow f). \label{F vs W}
\end{equation}
Letting the minimum of $F$ in Eq. (\ref{F vs W}) be equal to Eq. (3) in the main text, we have
\begin{equation}
F_{min} =-W_{max} = -U_{i}+U_{f} -Q_{max} =p \epsilon N - N H(p) k_{B}T \ln 2. \label{Fmin}
\end{equation}
Since $-U_{i}+U_{f} = p \epsilon N$ in our model, we arrive at the equivalence between the
thermodynamic and the information-theoretic entropies;
\begin{equation}
S_{th}= N H(p) k_{B} \ln 2. \label{thvsinf}
\end{equation}
It is interesting to see that both the thermodynamic and the information theoretic entropies are
defined as a limit, while the former is physically realized in the quasi-static limit and the
latter is achieved by the optimal limit of the data compression. The quasi-static processes means
that the operation has to be slow enough compared with microscopic processes. This can be viewed
as an aspect of the Markovian process, in which the information on the earlier configuration is
lost due to, e.g., the multiple-scattering of particles by the cylinder wall. The corresponding
process in our model is the erasure of information, which occurs when the partition in the memory
cell is removed. To lose the information stored in the memory, we need to wait for a while until
it becomes impossible to infer the history of molecule's trajectory.

As a byproduct, we can see that the extensivity of the thermodynamic entropy follows from that of
the Shannon entropy. Note also that Eq. (\ref{thvsinf}) can be viewed as the Boltzmann formula
$S_{th}=k_{B} \ln W$, where $W=2^{NH(p)}$ is the number of possible states under a given total
energy.

%%%%%%%%%%%%%%%%%%%%%%%%%%%%%%%%

\end{document}